\documentclass[aps,two column,floatfix,showpacs]{revtex4}
\usepackage{graphicx, bm}
\usepackage{amsmath}
\usepackage{amssymb}
\usepackage{color}

\begin{document}
\title{Three fermions in a box at the unitary limit: 
universality in a lattice model}
\author{L. Pricoupenko$^{1}$ and Y. Castin$^{2}$}
\affiliation
{$^{1}$Laboratoire de Physique Th\'{e}orique de la Mati\`{e}re Condens\'{e}e, 
Universit\'{e} Pierre et Marie Curie, case courier 121, 
4 place Jussieu, 
75252 Paris Cedex 05, France. \\
$^{2}$Laboratoire Kastler Brossel, Ecole normale sup\'erieure, UPMC, CNRS,
24 rue Lhomond, 75231 Paris Cedex 05, France. }
\date{\today}
\begin{abstract}
We consider three fermions with two spin components interacting
on a lattice model with an infinite scattering length.
Low lying eigenenergies in a cubic box with periodic boundary conditions,
and for a zero total momentum, are calculated numerically for decreasing 
values of the lattice period. The results are compared to the predictions 
of the zero range Bethe-Peierls model in continuous space, where the 
interaction is replaced by contact conditions. The numerical computation, 
combined with analytical arguments, shows the absence of negative energy
solution, and a rapid convergence of the lattice model towards the Bethe-Peierls 
model for a vanishing lattice period. This establishes for this system the 
universality of the zero interaction range limit.
\end{abstract}
\pacs{03. 75. Ss, 05. 30. Fk, 21. 45. +v}
\maketitle

Recent experimental progress has allowed to prepare a two-component Fermi
atomic gas in the BEC-BCS crossover regime and to study in the lab
many of its physical 
properties, such as the equation of state
of the gas and other thermodynamic properties, 
the fraction of condensed particles, the gap in the excitation
spectrum corresponding to the breaking of a pair, the superfluid properties
and the formation of a vortex lattice, the effect of a population
imbalance in the two spin-components and the corresponding possible quantum 
phases, $\ldots$
\cite{Thomas1, Thomas2, Jin,Salomon1,Jochim,Greiner,Zwierlein, Bourdel, 
Ketterle_vortex,Bartenstein,Grimm_gap, Hulet_xi, Ketterle_unba,
Stewart, Altmeyer, Luo}. 

The key to this impressive sequence of experimental 
results is the use of Feshbach
resonances \cite{Feshbach}: an external magnetic field ($B$) permits to
tune the two-body $s$-wave scattering length $a$ almost at will,
to positive or negative values, so that one can e.g.\ adiabatically
transform a weakly attractive Fermi gas into a Bose condensate of molecules. 
Interestingly, close to the resonance, the scattering length
diverges as $a\propto-1/(B-B_0)$ so that the infinite scattering
length regime ($|a|=\infty$) can be achieved. 
When the typical relative
momentum $k$ of the particles further satisfies $k b \ll 1, k|r_e| \ll 1$,
where $b$ is the range and $r_e$ the effective range of the interaction potential,
the $s$-wave scattering amplitude between two particles takes the maximal
modulus value $f_k= -1/(ik)$: This is the so-called unitary regime,
where the gas is strongly, and presumably maximally, interacting.

The unitary regime is achieved in present experiments for broad 
Feshbach resonances, that is for resonances where the effective range $r_e$
is of the order of the Van der Waals 
range of the interatomic forces \cite{Verhaar1, Verhaar2}. 
Examples of $s$-wave broad 
resonances are given for $^6$Li atoms by the one at $B_0 \simeq 830$~G
\cite{Thomas1, Jochim,Zwierlein, Salomon1} or also for $^{40}$K atoms at
$B_0 \simeq 200$~G \cite{Jin}. On a more theoretical point of
view, the unitary regime has the striking
property of being universal: E.g., the zero temperature equation
of state involves only $\hbar$, the atomic mass $m$, the atomic
density and a numerical constant independent of the atomic
species; this was checked experimentally,
this also appears in fixed node Monte Carlo simulations \cite{Pandharipande, Giorgini}
and more recently in exact Quantum Monte Carlo calculations 
\cite{Svistunov,Bulgac,Juillet}.  

In Refs.\cite{Svistunov,Juillet}, exact Quantum Monte Carlo simulations at 
the unitary regime are performed using a Hubbard model. From the condensed
matter physics point of view, this 
modelling of the system is a clever way to
avoid the fermionic sign problem. But it is more than
a theoretical trick in the case of ultra-cold atoms 
since it can be achieved experimentally by trapping atoms at the
nodes of an optical lattice in the tight-binding regime
\cite{Bloch}. 
The Bethe-Peierls zero range model is another commonly used way of modelling  
the unitary regime: pairwise interactions are replaced by contact 
conditions imposed on the many-body wave function 
\cite{BethePeierls,Efimov,Albeverio,Petrov3,
Petrov4,ShinaTan,Werner,YvanVarenna}. 
This model is very well adapted to analytical calculations in 
few-body problems \cite{Efimov,Werner}
but can also be useful to predict many-body properties
like time-dependent scaling solution \cite{Castinsca}, the link between short
range scaling properties and the energy of the trapped gas \cite{ShinaTan}, and
hidden symmetry properties \cite{hiddensym} of the trapped gas.

However, 
there is to our knowledge no general rigorous result
concerning the equivalence between the discrete (Hubbard model) 
and the continuous 
(Bethe-Peierls)  models for the unitary gas. 
As a crucial example, one may wonder 
if there is any few- or many-body bound state 
in a discrete model at the infinite scattering length limit. 
This is a non-trivial
question, since the infinite scattering length corresponds to
an attractive on-site interaction in the discrete model.

In this paper, we address this question for two and three fermions in a 
cubic box with periodic boundary conditions, 
when the interaction range tends to zero for a fixed infinite
value of the scattering length. 
Our results for the equivalence of the lattice model
and the Bethe-Peierls approach are analytical for two fermions but
still rely on a numerical step for three fermions.
In this few body problem, the grid spacing can however 
be made very small in comparison
to the grids currently used in Quantum Monte Carlo many-body calculations, 
thus allowing a more precise study of the zero lattice step limit and
a test of the linear scaling of thermodynamic quantities with
the grid spacing used in \cite{Svistunov}.
Our computations also
exemplify the remarkable property that short range physics 
of the binary interaction 
does not play any significant role in the unitary two-component Fermi gas,
and the fact that the Bethe-Peierls model is well behaved for three equal
mass fermions.

Our model is the lattice model used in the 
Quantum Monte Carlo simulations of
\cite{Juillet}.
It has already been described in details in Refs.\ \cite{Mora, QGLD2003}
so that we recall here only its main features. The
positions $\mathbf{r}_i$ of each particle $i$ are discretized on a
cubic lattice of period $b$. The Hamiltonian contains the kinetic
term of each particle, $\mathbf{p}^2/2m$,
such that the plane wave of wave vector
$\mathbf{k}$ has an energy
\begin{equation}
\epsilon_\mathbf{k} = \frac{\hbar^2 k^2}{2m}. 
\end{equation}
Here the wave vector is restricted to the first Brillouin zone of the
 lattice : 
\begin{equation}
\mathbf{k} \in {\mathcal D} \equiv [-\pi/b, \pi/b[^3. 
\end{equation}
We enclose the
system in a cubic box of size $L$ with periodic boundary conditions, 
so that the components $\{k_\alpha\}_{\alpha \in \{x, y, z\}}$ 
of $\mathbf{k}$ are
integer multiples of ${2\pi/L}$. In what follows we shall 
for convenience restrict 
our computations to the case where the ratio ${L/b=2N+1}$
is an odd integer, so that ${k_\alpha=2\pi n_\alpha/L}$ with ${n_\alpha \in
\{-N, -N+1, \hdots, N\}}$. The Hamiltonian also contains the interaction
potential between opposite spin fermions $i$ and $j$, which is a
discrete delta on the lattice:
\begin{equation}
V(\mathbf{r}_i, \mathbf{r}_j) = \frac{g_0}{b^3} \delta_{\mathbf{r}_i, \mathbf{r}_j}. 
\end{equation}
In \cite{QGLD2003} the 
matrix elements of the two-body $T$-matrix $\langle\mathbf{k}|T(E+i0^+)
|\mathbf{k}'\rangle$ for an infinite box size are shown to depend only on the
energy $E$, not on the plane wave momenta, which 
would imply in a continuous space a pure $s$-wave scattering.
The bare coupling constant $g_0$ is then adjusted in order to reproduce in the zero
energy limit the
desired value of the $s$-wave scattering length $a$ between two opposite spin
particles \cite{Mora, QGLD2003, YvanVarenna}:
\begin{equation}
\frac{1}{g_0}- \frac{1}{g} = -\int_{\mathcal{D}} \frac{d^3\mathbf{k}}{(2\pi)^3} \, \frac{1}{2\epsilon_\mathbf{k}} 
=- \frac{m K}{4\pi \hbar^2 b} , 
\label{eq:renormalize}
\end{equation}
where 
\begin{equation}
K=\frac{12}{\pi}\int_0^{\pi/4} d\theta\, \ln(1+1/\cos^2\theta)=2.442749\hdots
\end{equation}
may be expressed in terms of the dilog function,
and ${g=4\pi\hbar^2a/m}$ is the usual effective
$s$-wave coupling constant.
From the calculated energy dependence of the $T$-matrix, one may also extract 
the effective range $r_e$ of the interaction in the 
lattice model; $r_e$ is found to be proportional to the lattice period, 
$r_e\simeq 0.337\, b$ \cite{YvanVarenna}, and the limit
$b\to 0$ is equivalent to the limit of both zero range and zero effective 
range for the interaction \cite{two_channel}. As mentioned in the introduction,
this is the desired situation to reach the unitary limit when $|a|=\infty$.

We first solve the problem for two opposite spin fermions in the box, 
in the singlet spin state $|s\rangle= (|\uparrow\downarrow\rangle-
|\downarrow\uparrow\rangle)/\sqrt{2}$, 
by looking for eigenstates of eigenenergy 
$E$ with a ket of the form $|s\rangle\otimes|\phi\rangle$.
We restrict to the case of a zero total momentum \cite{et_donc},
so that the orbital part $|\phi\rangle$ may be expanded on
${|\mathbf{k}, -\mathbf{k}\rangle}=|1:\mathbf{k}\rangle\otimes
|2:-\mathbf{k}\rangle$, where $|1:\mathbf{k}\rangle$
is the normalized ket representing particle 1 with wave vector
$\mathbf{k}$. 
The corresponding wavefunction is
$\langle \mathbf{r}| \mathbf{k}\rangle = e^{i\mathbf{k}\cdot\mathbf{r}}/L^{3/2}$.
Schr\"{o}dinger's equation then reduces to:
\begin{equation}
(2 \epsilon_{\mathbf{k}} -E) \langle\mathbf{k}, -\mathbf{k}|\phi\rangle
+\frac{g_0}{L^{3/2}} \langle \mathbf{r}, \mathbf{r}|\phi\rangle=0, 
\label{eq:S2}
\end{equation}
where the last term does not depend on a common position $\mathbf{r}$
of the two particles. 
A first type of eigenstates corresponds to ${\langle \mathbf{r}, \mathbf{r}|
\phi \rangle = 0}$: these eigenstates have a zero probability to have two 
particles at the same point, and are also eigenstates of the
non-interacting case. An example of such a state with the correct
exchange symmetry is given by the wavefunction:
\begin{equation}
\phi(\mathbf{r}_1, \mathbf{r}_2) \propto 
\cos\left[\frac{2\pi}{L}(x_1-x_2)\right]-\cos\left[\frac{2\pi}{L}(y_1-y_2)\right] . 
\end{equation}

\begin{figure}[h]
\includegraphics[width=8cm,clip]{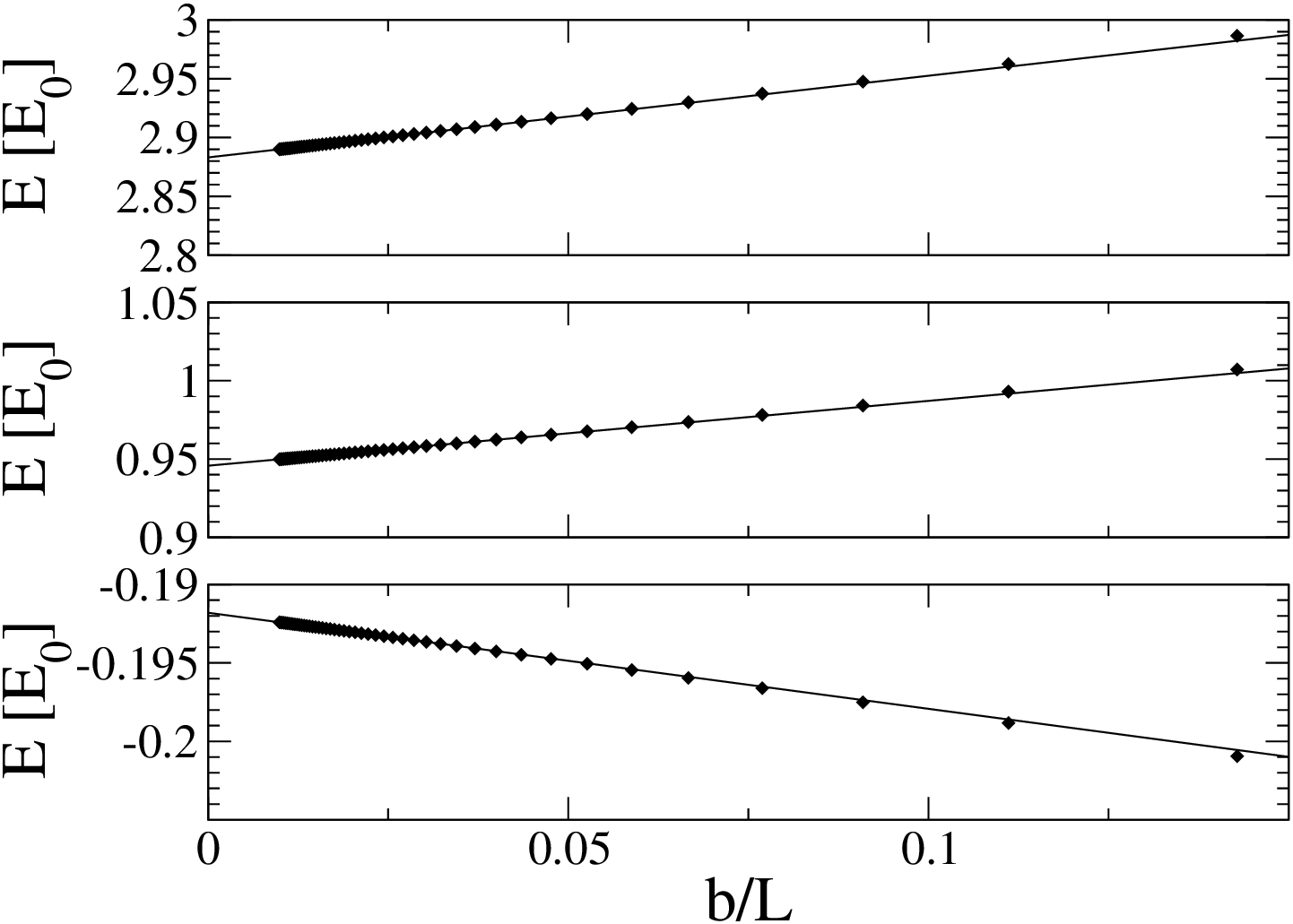}
\caption{First three eigenenergies for the interacting states
of two fermions in a box
of size $L$ for an infinite scattering length in the lattice model, as functions
of the lattice period $b$. The total momentum of the eigenstates
is fixed to zero. The computed eigenenergies are given by the plotting symbols,
in units of $E_0=(2\pi\hbar)^2/2mL^2$; the straight lines are linear fits performed
on the data with $b/L<2\times 10^{-2}$.}
\label{fig:2corps}
\end{figure}

We are interested here in the states of the second type, what we call
`interacting' states, such 
that ${\langle\mathbf{r}, \mathbf{r}|\phi\rangle\neq 0}$. 
Treating the interacting term in Eq.(\ref{eq:S2})
as a source term, one expresses $|\phi\rangle$ in terms of ${\langle
\mathbf{r}, \mathbf{r}|\phi\rangle}$ and a sum over $\mathbf{k}$. Projecting
the resulting expression onto ${|\mathbf{r}, \mathbf{r}\rangle}$ leads to a
closed equation (now ${E\neq 2\epsilon_\mathbf{k}}$):
\begin{equation}
\frac{1}{g_0}+ \frac{1}{L^3} \sum_{\mathbf{k}\in \mathcal{D}} 
\frac{1}{2\epsilon_\mathbf{k}-E}=0. 
\label{eq:eq_impl}
\end{equation}
The resulting implicit equation for $E$, of the form $u(E)=0$, where $u(E)$
is the left hand side of Eq.(\ref{eq:eq_impl}), is then readily solved
numerically; to this end, one notes that $u(E)$ has poles in each $E=2\epsilon_\mathbf{k}$,
and that it varies monotonically from $-\infty$ to $+\infty$
between two poles, so that $u(E)$ vanishes once and only once
between two successive values of $2\epsilon_\mathbf{k}$.
In Fig.\ref{fig:2corps}, we show for $|a|=\infty$ the first low lying
eigenenergies as functions of the lattice spacing; one observes a
convergence to finite values in the ${b/L\to 0}$ limit, with a first
correction scaling as ${b/L}$. 
A rewriting of the implicit equation for $E$ that will reveal
convenient in the ${b=0}$ limit is:
\begin{equation}
\frac{\pi L}{a} = \frac{(2\pi\hbar)^2}{m L^2} \left[\frac{1}{E} + 
\sum_{\mathbf{k}\in\mathcal{D}-\mathbf{0} }
\left(\frac{1}{E-2\epsilon_k}+\frac{1}{2\epsilon_k}\right)\right] + C(b) , 
\label{eq:ji}
\end{equation}
where the function $C(b)$ is defined by:
\begin{equation}
C(b) = \frac{(2\pi\hbar)^2 L}{2 m} \left( \int_{\mathcal D}\frac{d^3\mathbf{k}}{(2\pi)^3} \, \frac{1}{\epsilon_\mathbf{k}}
-\frac{1}{L^3} \sum_{\mathbf{k}\in \mathcal{D}-\mathbf{0}} \frac{1}{\epsilon_\mathbf{k}} \right), 
\label{eq:C(l)}
\end{equation}
and has a finite limit for ${b \to 0}$ which is given by $C(0) \simeq 8. 91364. $

We now briefly check that the $b=0$ limit in Eq.(\ref{eq:ji})
coincides with the prediction of the Bethe-Peierls model, which is a
continuous space model where one replaces the interaction potential by
the following contact conditions on the wavefunction \cite{BethePeierls,
Efimov,Albeverio,Petrov3,Petrov4,ShinaTan,Werner,YvanVarenna}: 
there exists a
function $S(\mathbf{R})$ such that, 
\begin{equation}
\phi(\mathbf{r}_1, \mathbf{r}_2)=
S(\mathbf{R})\left(\frac{1}{r}-\frac{1}{a} \right) + O(r) , 
\end{equation}
where ${r=|\mathbf{r}_1-\mathbf{r}_2|\to0}$ is the distance between
the two particles and the center of mass position
${\mathbf{R}=(\mathbf{r}_1+\mathbf{r}_2)/2}$ is fixed. At positions
$\mathbf{r}_1\neq\mathbf{r}_2$, the wavefunction solves
the free Schr\"odinger equation.
Using this model
we arrive at an implicit equation for the energy of an interacting
state exactly of the form obtained by taking the $b=0$ limit in
Eq.(\ref{eq:ji}), except that the constant $C(0)$ in the right hand
side is replaced by \cite{usual}:
\begin{equation}
C_{\rm BP} = \lim_{\sigma \to 0} \left(\int\! d^3\mathbf{u} \, \frac{e^{-u^2 \sigma^2}}{u^2} -\sum_{\mathbf{n}\in \mathbb{Z}^{3*}} \frac{e^{-n^2 \sigma^2}}{n^2}
\right). 
\label{eq:CBP}
\end{equation}
We expect the identity ${C_{\rm BP} =C(0)}$ from the general result
that the Bethe-Peierls model for the two-body problem reproduces the
zero range limit of a true interaction potential \cite{Albeverio,Olshanii}. It
is however instructive to check this property explicitly for the
 lattice model. One can show that:
\begin{equation}
C_{\rm BP} - C(0) =\lim_{\sigma \to 0} \sum_{\mathbf{n} \in \mathbb{Z}^{3*}} 
\int_{\mathcal I} d^3\mathbf{ u} \, \left[h_{\sigma}(\mathbf{n}+\mathbf{u})-h_{\sigma}(\mathbf{n})\right]
\end{equation}
where ${h_{\sigma}(\mathbf{q})=[\exp(-q^2 \sigma^2)-1]/q^2}$ and the
integration domain is ${{\mathcal I}=[-1/2, 1/2]^3} $. The desired
identity $C(0)=C_{\rm BP}$ results from the fact that one can exchange
the $\sigma=0$ limit and the summation over $\mathbf{n}$ in the above
equation \cite{justif, disagree}. 

In the lattice model, it is possible to show analytically
that the spectrum of the
 two-body problem for an infinite scattering length is bounded from below in the ${b\to 0}$
limit. Since ${g_0<0}$ for ${|a|=\infty}$, 
there exists at least one non-positive energy solution, by a variational argument. 
One then notes that the right hand side in Eq.(\ref{eq:ji}) is a strictly
decreasing function of $E$ over ${]-\infty, 0[}$, that tends to $-\infty$
in $E=0^-$, so that at most one negative
energy solution may exist. Furthermore one can show that the ${b \to 0}$
limit of the right hand side tends to $+\infty$ when ${E\to -\infty}$
\cite{minor}, whence this negative energy solution is finite \cite{neg_ener}. 

We now turn to the problem of three interacting fermions in the box. 
Schr\"{o}dinger's equation is obtained without loss of
generality by considering the particular
spin component ${(1\!\!:\, \uparrow\, ;2\!\!:\, \uparrow\, ;3\!\!:\, \downarrow)}$, so that the
interaction takes place only among the pairs ${(1, 3)}$ and ${(2, 3)}$, and
in the lattice model one obtains:
\begin{equation}
\left[
\sum_{i=1}^{3} \frac{\mathbf{p}_i^2}{2m} +\frac{g_0}{b^3} \left(\delta_{\mathbf{r}_1, \mathbf{r}_3} +\delta_{\mathbf{r}_2, \mathbf{r}_3}\right)- E 
\right]
\psi(\mathbf{r}_1, \mathbf{r}_2, \mathbf{r}_3) = 0. 
\label{eq:S3}
\end{equation}
We restrict to a zero total momentum modulo ${2\pi/b}$ along each
direction \cite{et_donc}; using the fermionic antisymmetry condition for the
transposition of particles 1 and 2, we express the part of 
Eq.(\ref{eq:S3}) involving the interaction in
terms of a function of the position of a single particle:
\begin{eqnarray}
\label{eq:def_f}
\psi(\mathbf{r}_1, \mathbf{r}_2, \mathbf{r}_1) &=& f(\mathbf{r}_2-\mathbf{r}_1) \\
\psi(\mathbf{r}_1, \mathbf{r}_2, \mathbf{r}_2) &=& -f(\mathbf{r}_1-\mathbf{r}_2). 
\end{eqnarray}
We then project Eq.(\ref{eq:S3}) on the plane waves in the box, which leads to:
\begin{equation}
\langle \mathbf{k}_1, \mathbf{k}_2, \mathbf{k}_3| \psi\rangle
=\frac{ g_0\, \delta^{\rm mod}_{\mathbf{k}_1+\mathbf{k}_2+\mathbf{k}_3, \mathbf{0}}}{
E-\epsilon_{\mathbf{ k}_1} -\epsilon_{\mathbf{ k}_2}-\epsilon_{\mathbf{ k}_3}}
\left(f_{\mathbf{k}_2} - f_{\mathbf{k}_1}\right)
\end{equation}
where $\delta^{\rm mod}$ is a discrete delta modulo $2\pi/b$ along each direction, and 
where the Fourier transform of $f(\mathbf{r})$ is defined as:
\begin{equation}
f_{\mathbf{k}} = \langle \mathbf{k}| f\rangle = \frac{b^3}{L^{3/2}} \sum_{\mathbf{r}
\in [0,L[^3}
\exp\left(-i\mathbf{k}\cdot\mathbf{r}\right) f(\mathbf{r}). 
\end{equation}
Replacing $f(\mathbf{r})$ in the right-hand side of this equation by
its expression in terms of $\langle\mathbf{k}_1, \mathbf{k}_2, \mathbf{k}_3|
\psi\rangle$ deduced from Eq.(\ref{eq:def_f}), 
we obtain a closed equation for $f_\mathbf{k}$:
\begin{equation}
\frac{L^3}{g_0} f_\mathbf{k} = f_\mathbf{k}\sum_{\mathbf{q} \in \mathcal D} a_{\mathbf{k}, \mathbf{q}}
-\sum_{\mathbf{q} \in \mathcal D} a_{\mathbf{k}, \mathbf{q}} f_\mathbf{q}
\label{eq:ls}
\end{equation}
where we have introduced the matrix:
\begin{equation}
a_{\mathbf{k}, \mathbf{q}} = \frac{1}{E-\epsilon_\mathbf{ k}-\epsilon_\mathbf{ q} - \epsilon_{[\mathbf{k}+\mathbf{q}]_{\mathrm{FBZ}}}} , 
\label{eq:mat}
\end{equation}
and for an arbitrary wavevector $\mathbf{u}$, 
$[\mathbf{u}]_{\mathrm{FBZ}}$ denotes the vector in the first
Brillouin zone that differs from $\mathbf{u}$ by integer multiples of
$2\pi/b$ along each direction. The eigenvalues $E$ of the three-body
problem are such that the linear system (\ref{eq:ls}) admits a
non-identically vanishing solution $f_\mathbf{k}$, that is the
determinant of this linear system is zero. Note that from
Eq.(\ref{eq:ls}), one has $f(\mathbf{0}) \propto \sum_{\mathbf{q} \in D}
f_\mathbf{k}=0$, a consequence of Pauli exclusion principle. 

\begin{figure}[h]
\includegraphics[width=8cm,clip]{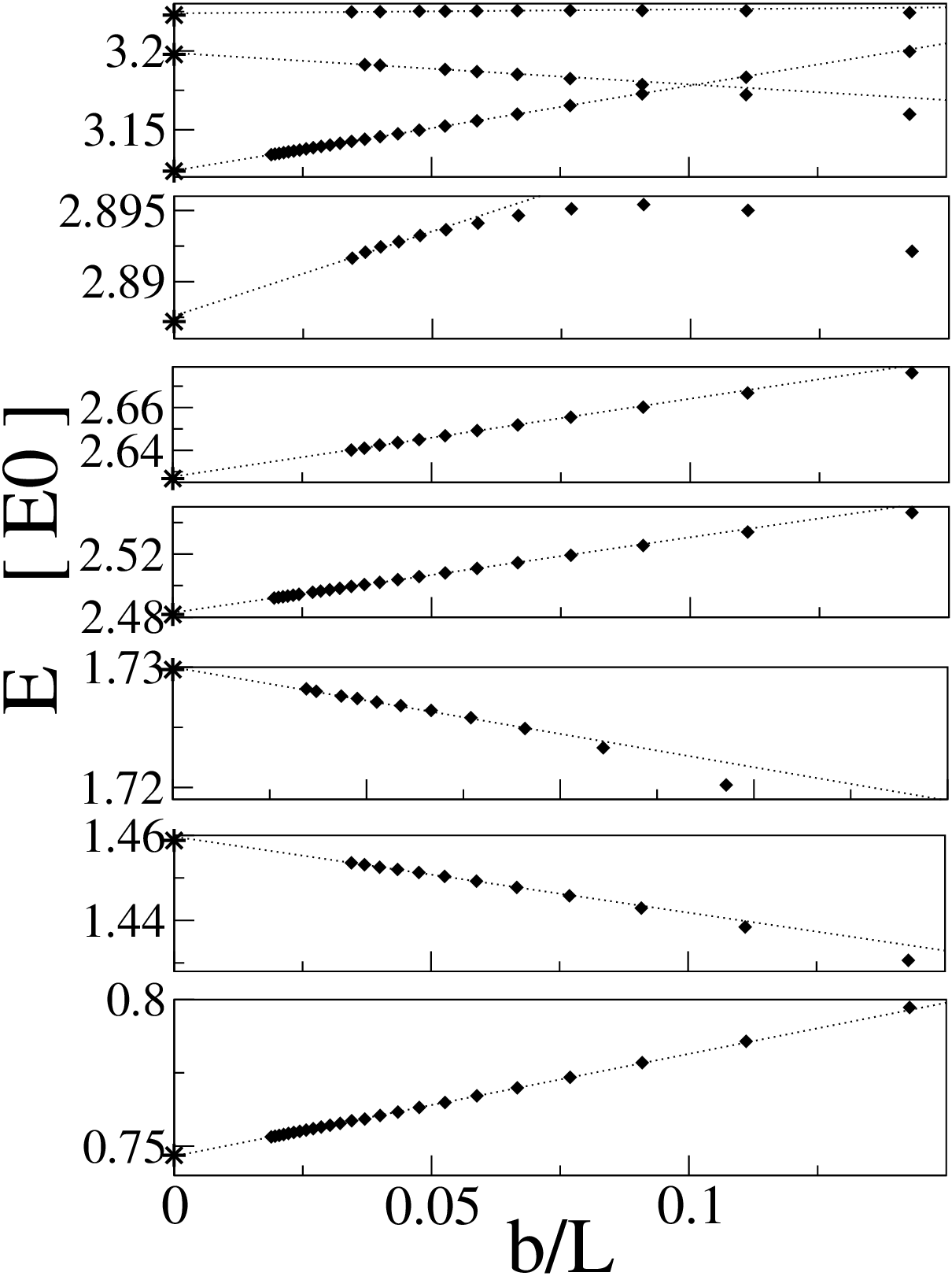}
\caption{
First eigenenergies of three fermions in a box of size $L$ for an infinite
scattering length in the lattice model, for a zero total momentum. 
The computed eigenenergies (diamonds) are given in units of 
${E_0=(2\pi\hbar)^2/2mL^2}$ for different values of
the lattice period $b$.  
For functions $f(\mathbf{r})$ invariant by reflection
along ${x,y,z}$ and by arbitrary permutation of ${x,y,z}$
we have computed the eigenenergies down to smaller values of $b/L$.
The straight lines are a linear fit
performed on the data over the range ${b/L\leq 1/15}$,
except for the energy branch $E\simeq 2.89 E_0$ which becomes more slowly
linear than the other branches.
The eigenenergies predicted by the Bethe-Peierls model
are given by stars in ${b=0}$.}
\label{fig:branches}
\end{figure}

For $|a| =\infty$, we have computed numerically
 the first
eigenenergies of the system, by calculating the determinant
as a function of $E$. In Fig.\ref{fig:branches} we give
these eigenenergies as functions of the ratio $b/L$. A rapid
convergence in the zero-$b$ limit is observed, with a linear
dependence in $b/L$. 

This rapid convergence illustrates the fact that equal mass fermions
easily exhibit universal properties, as revealed by experiments; here
$b$ plays the role of the finite Van der Waals range of the true
potential [given by $(m C_6/\hbar^2)^{1/4}$, where $C_6$ is the Van der
Waals coefficient], and $L$ is of the order of the mean interparticle
distance in a real gas. As an example, for $^6$Li atoms $b \sim 3$~nm
and in experiments for the broad Feshbach resonance in the
$s$-wave channel at $\sim 830$~G the atomic density is of the order of
$10^{13}$~cm$^{-3}$, so that the ratio $b/L$ is of the order of
$10^{-2}$ which is well within the zero-$b$ limit. 

The absence of negative three-body eigenenergies in the unitary limit can be obtained
numerically very efficiently
through a formal analogy between Eq.(\ref{eq:ls}) 
and a set of rate equations on fictitious occupation
numbers of the single particle modes in the box. Assuming $E\leq 0$,
we note $\Pi_{\mathbf{ k}}$ the
fictitious occupation number in the mode
$\mathbf{k}$ and $\Gamma_{\mathbf{k} \to
\mathbf{q}}=g_0 a_{{\mathbf q}, {\mathbf k} }/L^3$ the transition rate
from the mode $\mathbf{k}$ to the mode $\mathbf{q}$. From
Eq.(\ref{eq:mat}), one obtains the property $\Gamma_{\mathbf{k} \to
\mathbf{q}}= \Gamma_{\mathbf{q} \to \mathbf{k}}$, and the rate equation can
be written as:
\begin{equation}
\frac{d \Pi_{\mathbf k}}{dt} = -\left( \sum_{\mathbf{q}\neq \mathbf{k}}\Gamma_{\mathbf{k} \to \mathbf{q}} \right) \Pi_{\mathbf k} + \sum_{\mathbf{q}\neq \mathbf{k}}\Gamma_{\mathbf{q} \to \mathbf{k}}\Pi_{\mathbf q}
\label{eq:robinet} . 
\end{equation}
The symmetric matrix $M(E)$, which defines the first order linear system in
Eq.(\ref{eq:robinet}), $d\mbox{\boldmath$\Pi$}/dt=M(E) \mbox{\boldmath$\Pi$}$, 
has the following properties: 1) its
eigenvalues are non-positive, since it is a set of rate equations; 
2) its eigenvalues are decreasing
function of the energy $E$, which can be deduced from the fact that
$dM(E)/dE$ is a matrix of rate equations and obeys property 1), and
from the Hellman-Feynman theorem;
and 3) eigenmodes of Eq.(\ref{eq:robinet})
with an eigenvalue equal to~$-1$ correspond to solutions
$f_{\mathbf{k}}$ of Eq.(\ref{eq:ls}) with ${\Pi_{\mathbf
k}=f_{\mathbf{k}} \exp(-t)}$. Therefore, in order to check that there is
no non-zero solution of Eq.(\ref{eq:ls}) for $E<0$, it is sufficient to
check that all eigenvalues of $M(E=0)$ are strictly larger than $-1$. 

\begin{figure}[h]
\includegraphics[width=8cm,clip]{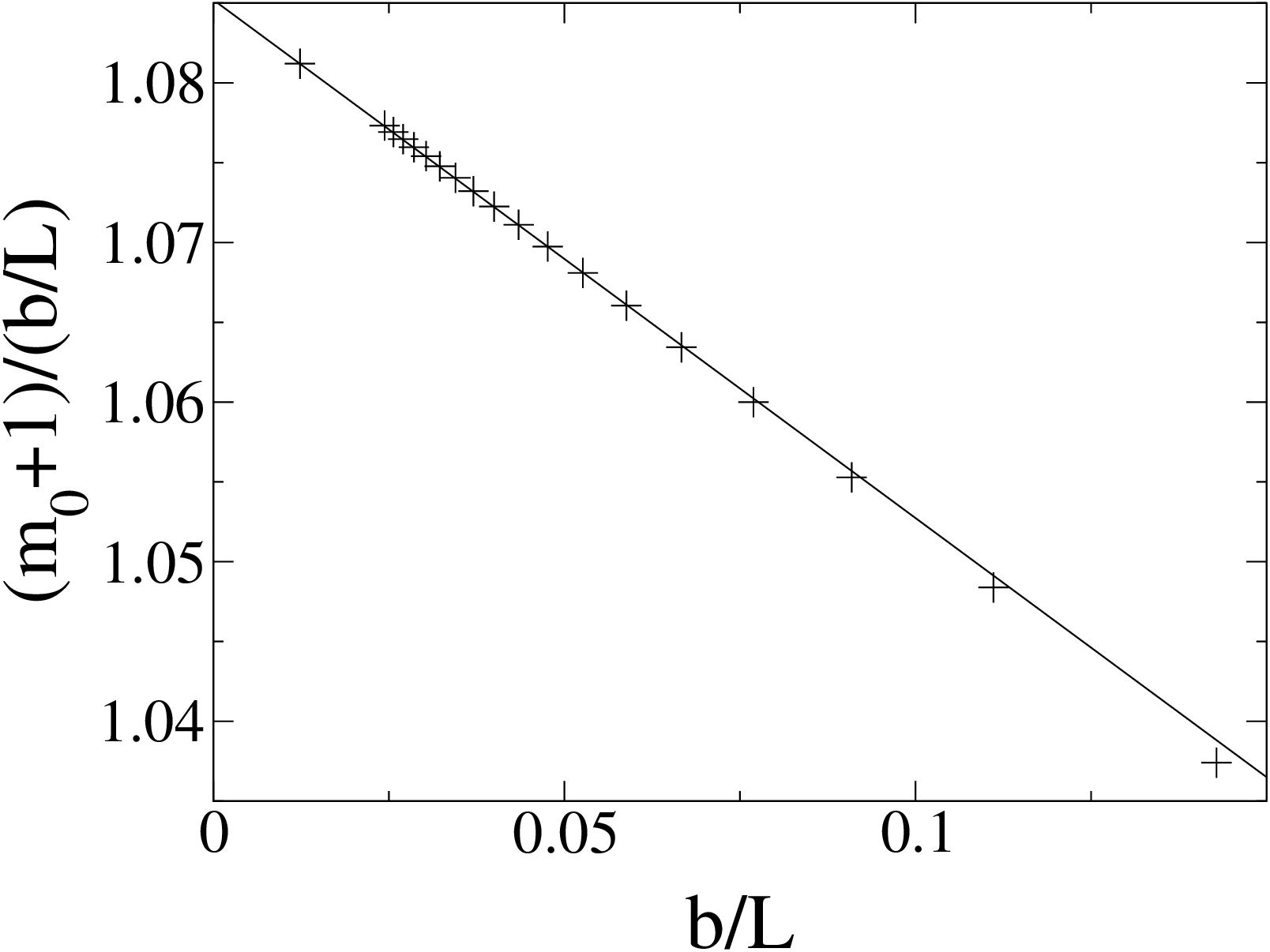}
\caption{
Quantity $(m_0+1)/(b/L)$ as a function of the lattice period $b$.
Here $m_0$ is the lowest eigenvalue of the matrix $M(E)$ defining the linear
system Eq.(\ref{eq:robinet}), for
$E=0$ and for an infinite scattering length.
The fact that $m_0+1>0$ shows that there is no negative
eigenenergy for the three fermions, see text. The symbols are obtained from
a numerical calculation of $m_0$.
The solid line is a linear fit over the range $b/L\leq 1/29$, 
not including the point with $b/L=1/81$: for this point, the matrix
$M$ has more than half a million lines so that $m_0$ was obtained
by a computer memory-saving iterative method rather
than by a direct diagonalisation.}
\label{fig:m0}
\end{figure}

We have computed the lowest eigenvalue $m_0$ of the matrix $M(E=0)$
as a function of the ratio ${b/L}$.
A fit of $m_0$ as a
function of $b/L$ suggests $\lim_{b \to 0}m_0 \simeq -1$. To better see
what happens in the zero $b/L$ limit, we note that having $m_0>-1$ is equivalent
to having $(m_0+1)/g_0<0$, or more simply $(m_0+1)/(b/L)>0$. 
We have thus plotted in Fig.\ref{fig:m0} 
the ratio $(m_0+1)/(b/L)$, which is seen to tend 
to a positive value for $b\to 0$, 
$\simeq 1.085$, with a negative slope;  
this excludes
the existence of negative eigenenergies for the three fermions
at infinite scattering length
even in the small $b$ limit \cite{tbs}.

In a last step, we compare the results of the lattice model to the
predictions of the Bethe-Peierls approach for three fermions in a continuous space, 
which was shown 
to be a successful model in free space \cite{Petrov3,Petrov4} and in a harmonic trap 
at the unitary limit \cite{Werner}.
For this
purpose, we introduce the function $F_\mathbf{k}$ which is the Fourier
transform of the regular part of the wave function as
${|\mathbf{r}_1-\mathbf{r}_3|\to 0}$:
\begin{equation}
F(\mathbf{R})= \lim_{r\to 0} \left[r \psi\left(\mathbf{R}+\frac{\mathbf{r}}{2}, 
\mathbf{0}, \mathbf{R}-\frac{\mathbf{r}}{2}\right)\right], 
\end{equation}
where we have used the translational invariance. By reproducing a
calculation procedure analogous to what we have done for the lattice 
model, we obtain the following infinite dimension linear system:
\begin{eqnarray}
 \frac{L^3}{g} F_\mathbf{k} &=&F_\mathbf{k}
\left[A_{\mathbf{k}, \mathbf{0}} + \sum_{\mathbf{q}\neq\mathbf{0}}\left( A_{\mathbf{k}, \mathbf{q}} 
+ \frac{1}{2 \epsilon_\mathbf{ q}}\right) + \frac{mL^2C_{\mathrm{BP}}}{(2\pi\hbar)^2} \right]
\nonumber\\
&&\qquad -\sum_{\mathbf{q}}A_{\mathbf{k}, \mathbf{q}} F_\mathbf{q} \quad , 
\label{eq:3bodyBP}
\end{eqnarray}
where the wavevectors $\mathbf{k}$ and $\mathbf{q}$ now run over the whole
space $(2\pi/L) \mathbb{Z}^3$, and: 
\begin{equation}
A_{\mathbf{k}, \mathbf{q}}= \frac{1}{E-\epsilon_\mathbf{k} - \epsilon_\mathbf{q} - \epsilon_\mathbf{k+q} } \quad . 
\end{equation}

The similarity between the structure of (\ref{eq:ls}) and
(\ref{eq:3bodyBP}) is apparent. 
Numerically, at $|a|=\infty$,  we have verified the convergence
between the two models as $b \to 0$ in Eq.(\ref{eq:ls}), 
see Fig.\ref{fig:branches}.
Analytically, one can even formally check the equivalence 
between the two sets of
equations (\ref{eq:ls}) and (\ref{eq:3bodyBP}): First, we
eliminate the integral of $1/\epsilon_\mathbf{k}$ between (\ref{eq:renormalize})
and (\ref{eq:C(l)}), to express $1/g_0$ in terms of $1/g$ and $C(b)$. 
Second, we replace $1/g_0$ by the resulting expression in
Eq.(\ref{eq:ls}). Third, we take the limit $b\to 0$: we exactly
recover the system (\ref{eq:3bodyBP}). Hence, if the eigenenergy $E$
and the corresponding function $f$ in the lattice model
have a well defined limit for $b=0$, 
this shows that the limit is given by the Bethe-Peierls model. 
Of course, the real mathematical difficulty is to show the existence of the limit, 
in particular for all eigenenergies. This property is not granted:
For example, the present lattice model generalized to the case of a $\downarrow$ particle
of a mass $m_3$ different from the mass $m$ of the two $\uparrow$ particles
leads, for a large enough mass ratio $m/m_3$, to a three-body
energy spectrum not bounded from below in the $b=0$ limit, even
though the Pauli exclusion principle prevents from having the three
particles on the same lattice site \cite{varia}.

In conclusion, we have computed numerically the low lying eigenenergies 
of three spin-1/2 fermions in a box, interacting with an infinite
scattering length in a lattice model, for a zero total 
momentum and for
decreasing values of the lattice period. Our results show numerically
the equivalence between this model and the Bethe-Peierls approach in
the limit of zero lattice period. This is related to the fact 
that the eigenenergies $E$ are bounded from below in the zero lattice 
period limit $b\to 0$, more precisely $E>0$. 
Such a convergence of the eigenstates of fermions in a lattice model
towards universal states when $b\to 0$ is a key property 
used in Monte Carlo simulations at the $N$-body level
\cite{Svistunov,Bulgac,Juillet}. 

We thank F.\ Werner for interesting discussions on the subject. 
Laboratoire de Physique Th\'{e}orique de la Mati\`{e}re Condens\'{e}e
is the Unit\'{e} Mixte de Recherche 7600 of Centre National de la
Recherche Scientifique (CNRS).
% Laboratoire Kastler Brossel (LKB) is a research unit
%of Ecole normale sup\'{e}rieure and of Universit\'{e} Pierre et Marie
%Curie,  associated to CNRS.
The cold atom group at LKB is a member of IFRAF.


\begin{thebibliography}{99}
\bibitem{two_channel} 
In contrast, 
in a two-channel model for a Feshbach resonance, one finds that 
the effective range has a finite (and negative) limit 
in the zero potential range limit (the so-called narrow Feshbach
resonance limit) \cite{YvanVarenna}.
\bibitem{et_donc}
One thus cannot conclude that the corresponding minimal
eigenenergy is the absolute ground state energy.
\bibitem{usual}
Usually one expresses the Green's
function of the Laplacian in a cubic box in terms of plane
waves. This leads to ${C_{\rm BP}= \lim_{x\to 0} v(\mathbf{x})}$, with
$v(\mathbf{x})=\int\!d^3\mathbf{u}\, \exp(i\mathbf{u}\cdot\mathbf{
x})/u^2-\sum_{\mathbf{n}\in\mathbb{Z}^{3*}} \exp(i\mathbf{
n}\cdot\mathbf{x})/n^2$. This definition of ${v(\mathbf{x})}$ should be
understood within the frame of the theory of distributions. We define
the $x=0$ limit of $v(\mathbf{x})$ as the limit for ${\sigma\to 0}$ of ${\int\!d^3\mathbf{x}\, 
v(\mathbf{x}) \phi(\mathbf{x}/\sigma)/\sigma^3}$, 
where $\phi$ is a $C^\infty$ rapidly
decreasing function with ${\int\!d^3\mathbf{x}\, \phi(\mathbf{x})=1}$. In
Eq.(\ref{eq:CBP}) we have taken for simplicity $\phi$ to be a Gaussian, 
but we have shown that $C_{\rm BP}$ is independent of
this choice.
\bibitem{justif}
One uses the rewriting $h_\sigma(\mathbf{n}+\mathbf{u})-h_\sigma(\mathbf{n})=
T_1+T_2$, with $T_1=
[\hat{\phi}[\sigma(\mathbf{n}+\mathbf{u})]-\hat{\phi}(\sigma\mathbf{n})]/(\mathbf{n}
+\mathbf{u})^2$, $T_2=[\hat{\phi}(\sigma\mathbf{n})-1][1/(\mathbf{n}+\mathbf{u})^2
-1/n^2]$ and $\hat{\phi}(\mathbf{x})=\exp(-x^2)$.
Using a large $n$ expansion, one finds that the integral of $T_2$ over the
symmetric integration domain $\mathcal{I}$ is $O(1/n^4)$, so that the theorem
of dominated convergence applies.
For $T_1$, one uses the Taylor-Lagrange formula up to second order for
the numerator:
For a given $\mathbf{u}$,
there exists a vector $\mathbf{x}_\mathbf{u}$ on the line
connecting $\sigma \mathbf{n}$ and $\sigma (\mathbf{n}+\mathbf{u})$
such that
$\hat{\phi}[\sigma(\mathbf{n}+\mathbf{u})]-\hat{\phi}(\sigma\mathbf{n})=
\sum_i \sigma u_i \partial_i\hat\phi(\sigma\mathbf{n})
+\frac{1}{2} \sum_{i,j} \sigma^2 u_i u_j \partial_i\partial_j \hat{\phi}(\mathbf{x}_\mathbf{u}).$
The term involving the first order derivatives of $\hat{\phi}$
vanishes after integration over $\mathbf{u}$. Since the second order derivatives of 
$\hat{\phi}(\mathbf{x})$
are rapidly decreasing functions, they are in particular $\leq A/x^2$ at large
$x$ for some number $A$, so that  the integral of $T_1$ over $\mathcal{I}$ is 
bounded by $A/n^4$ and the theorem of dominated converge applies again.
\bibitem{minor}
One uses the fact that for $n \in \mathbb{N}^{*}$, ${\epsilon/[n^2(n^2+\epsilon)]}$ is
positive when ${\epsilon>0}$, and tends to $1/n^2$ for ${\epsilon\to +\infty}$. The fact
that ${\sum_{\mathbf{n}\in \mathbb{Z}^{3*}}1/n^2 =+\infty}$ gives the result.
\bibitem{neg_ener}
In the limit $b\to 0$, there exists a negative energy solution $E<0$ for
all $a$. Its energy can be calculated accurately directly from
the Bethe-Peierls model from a more convenient
representation of the function $v(\mathbf{x})$ in \cite{usual}, using Poisson's
summation formula applied to the function 
$\mathbf{u}\to e^{i\mathbf{u}\cdot\mathbf{x}}/(u^2+\lambda^2)$
where $\lambda>0$ is arbitrary. One obtains
$C_{\rm BP}=\lambda^{-2}+2\pi^2\lambda- \sum_{\mathbf{n}\in \mathbb{Z}^{3*}}
[\lambda^2 n^{-2} (\lambda^2+n^2)^{-1} + \pi \exp(-2\pi\lambda n)/n]$, an expression
whose value does not depend on $\lambda$. 
Specializing to the unitary limit, and taking $\lambda=\alpha/(2\pi)$,
where $\alpha=1.945766\ldots$ solves $\alpha =
\sum_{\mathbf{n}\in\mathbb{Z}^{3*}} \exp(-\alpha n)/n$, one finds a
minimal eigenenergy $E=-\alpha^2\hbar^2/m L^2$.
\bibitem{varia} 
For an arbitrary mass ratio,
the coupling constant $g_0$ for an infinite scattering length
is $g_0=-2\pi\hbar^2b/(\mu K)$ where $1/\mu = 1/m + 1/m_3$ is the inverse of
the reduced mass. One may take as a simple variational ansatz the ground state of the
three-body problem for $m=\infty$, of the form
$|\psi_\infty\rangle = [|\mathbf{r}_1\rangle |\mathbf{r}_2\rangle - |\mathbf{r}_2\rangle
|\mathbf{r}_1\rangle] |\chi\rangle$, with $\mathbf{r}_1-\mathbf{r}_2 = b\, \mathbf{e}_x$
and $\mathbf{e}_x$ the unit vector along $x$; $|\chi\rangle$ has a simple
expression in momentum space and one finds 
$\chi(\mathbf{r}_1)=\chi(\mathbf{r}_2)$. 
For a finite value of $m$ the expectation value of
$H$ in $|\psi_\infty\rangle$ gives an upper bound $E_v$ 
on the ground state three-body energy,
\[
E_v =  \frac{\hbar^2 \pi^2}{2 m_3 b^2} \left(A + B\frac{m_3}{m}\right)
\]
where $A$ is the smallest root of $F(A)=1+\int_{[-1,1]^3} d^3q
[1+\cos(\pi \mathbf{q}\cdot\mathbf{e}_x)]/[2 \pi K (A-q^2)]$
and $B=2+1/F'(A)$. One finds 
 $A\simeq -0.042088$ and $B\simeq 1.75762$. Then $E_v\to -\infty$ when
$b\to 0$ for a mass ratio  $m/m_3$ above the critical value $\simeq 41.8$. 
Actually the exact critical mass ratio is expected to be below $13.6069\ldots$ 
since the Efimov phenomenon takes place for $m/m_3>13.6069\ldots$ 
\cite{mass_ratio}.
{\sl Note}: in the bosonic case, for the lattice model at $|a|=\infty$ with $N_B$ bosons of mass $m$ in the
same spin state,
one may take as a variational ansatz the state vector 
where all the $N_B$ bosons are on the same
 lattice site; one then finds an upper bound on the ground state energy 
$E_v^B= g_0 N_B [N_B -(1+\pi K/4)] /(2 b^3)$, with $1+\pi K/4 \simeq 2.9185$,
so that the ground state energy tends to $-\infty$ for $b\to 0$ 
if $N_B \geq 3$.
\bibitem{mass_ratio}
 From \cite{Petrov3} we find that 
the minimal mass ratio $m/m_3$ leading to the Efimov phenomenon solves
$-\frac{\pi}{2} \sin^2(2\theta)+\cot 2\theta+2\theta=0$,
excluding the trivial root $\theta=\pi/4$,
with $\theta=\arctan[(1+2m/m_3)^{1/2}]$.
\end{thebibliography}
\end{document}